\documentclass[conference]{IEEEtran}
\IEEEoverridecommandlockouts
\usepackage{cite}
\usepackage{amsmath,amssymb,amsfonts}
\usepackage{algorithmic}
\usepackage{graphicx}
\usepackage{textcomp}
\usepackage{xcolor}
\usepackage{subfig}
\usepackage{booktabs}
\usepackage[hidelinks]{hyperref}
\def\BibTeX{{\rm B\kern-.05em{\sc i\kern-.025em b}\kern-.08em
    T\kern-.1667em\lower.7ex\hbox{E}\kern-.125emX}}
\usepackage[framemethod=tikz]{mdframed}
\mdfdefinestyle{mystyle}{%
  rightline=true,
  innerleftmargin=10,
  innerrightmargin=10,
  outerlinewidth=2pt,
  topline=false,
  rightline=true,
  bottomline=false,
  skipabove=\topsep,
  skipbelow=\topsep,
  frametitlefont=\normalfont\sc
}
\begin{document}
\begin{sloppy}

\title{Adding Context to Source Code Representations\\for Deep Learning}

\author{\IEEEauthorblockN{Fuwei Tian}
\IEEEauthorblockA{
\textit{The University of Melbourne}\\
Australia\\
fuweit@student.unimelb.edu.au}
\and
\IEEEauthorblockN{Christoph Treude}
\IEEEauthorblockA{
\textit{The University of Melbourne}\\
Australia\\
christoph.treude@unimelb.edu.au}
}

\maketitle

\begin{abstract}
Deep learning models have been successfully applied to a variety of software engineering tasks, such as code classification, summarisation, and bug and vulnerability detection. In order to apply deep learning to these tasks, source code needs to be represented in a format that is suitable for input into the deep learning model. Most approaches to representing source code, such as tokens, abstract syntax trees (ASTs), data flow graphs (DFGs), and control flow graphs (CFGs) only focus on the code itself and do not take into account additional context that could be useful for deep learning models. In this paper, we argue that it is beneficial for deep learning models to have access to additional contextual information about the code being analysed. We present preliminary evidence that encoding context from the call hierarchy along with information from the code itself can improve the performance of a state-of-the-art deep learning model for two software engineering tasks. We outline our research agenda for adding further contextual information to source code representations for deep learning.
\end{abstract}


\section{Introduction}

Program comprehension is a complex task that often requires developers to refer to multiple software artefacts~\cite{maletic2001supporting} which might be useful in helping developers construct a mental model of a program~\cite{latoza2006maintaining}. Experienced developers are better at deciding which cues from different artefacts might be useful to aid in program comprehension~\cite{kulkarni2014supporting}. For example, Kulkarni and Varma propose a developer's perception model for program investigation which indicates the role of twelve artefacts such as control flow, version history changes, and bug reports in comprehension tasks related to concepts, procedures, features, and modules~\cite{kulkarni2014supporting}. In another example, eye tracking has found that developers looked beyond a particular method to understand that method~\cite{sharif2016tracking}. This type of work has well established that developers use additional context when analysing source code---not just the code itself.

In this paper, we argue that when asking deep learning models to comprehend source code, we should give them the same benefit, i.e., access to information beyond the source code that the deep learning model is asked to analyse. The field of source code representation has seen many recent advances, with most work focusing on improving models by applying the latest innovations from the deep learning research community. Different data structures have been explored, including tokens, trees, and graphs~\cite{samoaa2022systematic}, to encode lexical, syntactic, and/or semantic information~\cite{jiang2022hierarchical}. However, the input to all these approaches has mostly remained the same---a snippet of source code.

Our preliminary work presented in this paper provides evidence that adding contextual information to the input of the state-of-the-art code representation approach ASTNN~\cite{zhang2019novel} can improve the performance of two software engineering tasks that are often used to evaluate the quality of code representation, i.e., clone detection~\cite{roy2009comparison} and code classification~\cite{ugurel2002s}. ASTNN takes code fragments as input and uses a neural network based on abstract syntax trees (ASTs) to capture statement-level lexical and syntactical knowledge as well as the naturalness of statements. We investigate the performance of ASTNN on source code methods from the SeSaMe dataset~\cite{kamp2019sesame} of semantically similar Java methods, using the publicly available implementation of ASTNN as a baseline and comparing it to modified versions which add context from a method's call hierarchy to the input. We explore different alternatives for encoding context and combining it with the encoding of a method into a combined representation.

Adding context from the call hierarchy (i.e., caller and callee context) can improve the performance in the clone detection scenario by 8\% (from an F1 score of 0.706 to an F1 score of 0.765). Interestingly, this performance improvement can only be achieved when we encode the difference between methods along with the information from their callers and callees, see Section~\ref{sec:aggregation} for details. Concatenation and max-pooling do not have a positive effect on performance in this scenario. In the code classification scenario, we observe performance improvements of 11\% for concatenation (from an F1 score of 0.633 to an F1 score of 0.704) and 5\% for max-pooling (from an F1 score of 0.711 to an F1 score of 0.747). Interestingly, performance gains differ between adding caller and/or callee context, and adding the callee context only actually decreased performance.

Adding context of source code fragments can improve the quality of source code representations for deep learning. However, which context to encode, how to encode it, and how to combine its encoding with the encoding of the original source code fragment all have implications on the performance of a deep learning model for downstream software engineering tasks. Based on these insights, we put forward our research agenda for adding context to source code representations for deep learning, with a focus on which context to add, how to combine source code fragments and their context, and dissecting how context plays a role in the deep learning models. We summarise related work in Section~\ref{sec:related} and present our preliminary study in Section~\ref{sec:study} before we conclude with our research agenda in Section~\ref{sec:agenda}.

\section{Related Work}
\label{sec:related}

Our work lies at the intersection of related work on source code representation and on the role of context in software engineering. 

\subsection{Source Code Representations}

In recent years, there has been a lot of work done on representing source code for machine learning applications. This can roughly be divided into work based on lexical, syntactical, and semantic information~\cite{jiang2022hierarchical}. In approaches aimed at representing \textit{lexical} information, a program is transformed into a sequence of tokens. For example, related work has shown that n-gram models can successfully handle token prediction across different project domains, given a large corpus for training~\cite{allamanis2013mining}, and n-gram models have been used to synthesise code completions for API method calls~\cite{raychev2014code}.

Approaches aimed at representing \textit{syntactical} information often rely on the AST. They use heuristic rules~\cite{jiang2007deckard} or machine and deep learning algorithms~\cite{zhang2019novel} for encoding information from the AST, e.g., in the form of a vector. For example, ASTNN~\cite{zhang2019novel} splits each large AST into a sequence of smaller sub-trees, and then learns syntactic knowledge from each sub-tree separately. Approaches aimed at representing \textit{semantic} information additionally incorporate code dependency information, often related to data flow and control flow information~\cite{zhao2018deepsim}. Recent work has shown that combining low-level syntactic information and high-level semantic information can improve source code representation for multiple program comprehension tasks~\cite{jiang2022hierarchical}. Hybrid representation approaches which combine mulitple representations are becoming increasingly common, compared to representations that use tokens, trees, or graphs only~\cite{samoaa2022systematic}.

None of these methods have focused on the input to source code representation. In this work, we argue that adding additional context from outside of the code fragment that is to be represented has the potential of improving the performance of any of the approaches. 

\subsection{Context in Software Engineering}

The need for context in software engineering is well established. For example, IDEs need context to understand the task they are supporting~\cite{murphy2018need}, developers need context to navigate technical discussions on Stack Overflow~\cite{galappaththi2022does}, and tools need context to automatically process source code~\cite{mcburney2015automatic, haque2020improved}. Context can include static artefacts such as documentation~\cite{dagenais2012recovering}, historical information such as past changes~\cite{zimmermann2005mining}, dynamic execution information such as traces~\cite{ko2004designing}, individual developer activity such as IDE interactions~\cite{kersten2006using}, and team and organisation activity such as communication and coordination archives~\cite{chatterjee2019exploratory}.

In this paper, we argue that all of these forms of context are potentially useful to augment the input to source code representation approaches---just as human developers have access to this information, deep learning models might benefit from this additional context. In our preliminary study (see next section), we rely on the call hierarchy for context, similar to the source code summarisation work by McBurney and McMillan~\cite{mcburney2015automatic}.

\section{Preliminary Study}
\label{sec:study}

In this section, we present our preliminary study to establish that adding context can indeed improve the performance of a state-of-the-art deep learning model for software engineering tasks. As a first step, we use the call hierarchy of a source code method as its context, but we believe that a similar methodology can be applied to other types of context, such as the ones described above.

\subsection{Research Questions}

To understand the potential of our idea and investigate how to best implement it, we ask two research questions:

\begin{description}
\item[RQ1] What is the impact of encoding additional context on the performance of a state-of-the-art deep learning model?
\item[RQ2] What is the impact of different approaches to aggregate the representation of code and its context?
\end{description}

\subsection{Data Collection}

To conduct our experiments, we chose the SeSaMe dataset which contains 857 Java method pairs from eleven open source projects that have been manually classified according to their semantic similarity~\cite{kamp2019sesame}. Unlike popular datasets such as BigCloneBench~\cite{svajlenko2015evaluating}, SeSaMe contains the repository link for each of its methods which allows us to extract their call hierarchies.\footnote{\url{https://github.com/gousiosg/java-callgraph}} For each method in the SeSaMe dataset, we extract its callers and callees. If a method has multiple callers and callees, we choose the largest one to encode as additional context. We use the state-of-the-art code representation approach ASTNN~\cite{zhang2019novel} as a baseline. Exploring other approaches as baselines, in particular those based on graph representations, is part of our future work. Since ASTNN relies on ASTs, we parse each method as well as its caller and callee methods into an AST. We exclude methods that cannot be parsed. 

\begin{table}
\centering
\caption{Datasets}
\label{tab:data}
\begin{tabular}{lr}
\toprule
Dataset                   & Count \\
\midrule
Classification Train Set  & 834   \\
Classification Dev Set    & 104   \\
Classification Test Set   & 105   \\
\midrule
Clone Detection Train Set & 448   \\
Clone Detection Dev Set   & 56    \\
Clone Detection Test Set  & 57    \\
\bottomrule
\end{tabular}
\end{table}

Following the long line of work on source code representations (e.g.,~\cite{zhang2019novel}), we use clone detection and source code classification as downstream software engineering tasks for evaluating the models:
\begin{itemize}
\item Code Clone Detection: For each method pair, the SeSaMe dataset contains up to eight valid ratings from expert programmers for semantic similarity in terms of \textit{goals}, \textit{operations}, and \textit{effects}, along with the corresponding \textit{confidence}. To construct our ground truth, we apply weights of 0.6, 0.8, and 1 to the confidence of low, medium, and high, respectively, and then average the data across valid ratings and dimensions to create a binary label.
\item Code Classification: We treat the origin of each source code method as its class, resulting in eleven classes since the SeSaMe dataset contains methods from eleven open source projects.
\end{itemize}

We adopt an 80\%, 10\%, 10\% split for training, validation, and testing. Table~\ref{tab:data} shows the size of the corresponding datasets.

\subsection{Aggregation}
\label{sec:aggregation}

\begin{figure*}
    \centering
    \subfloat[Concatenation]{\includegraphics[width=0.33\linewidth]{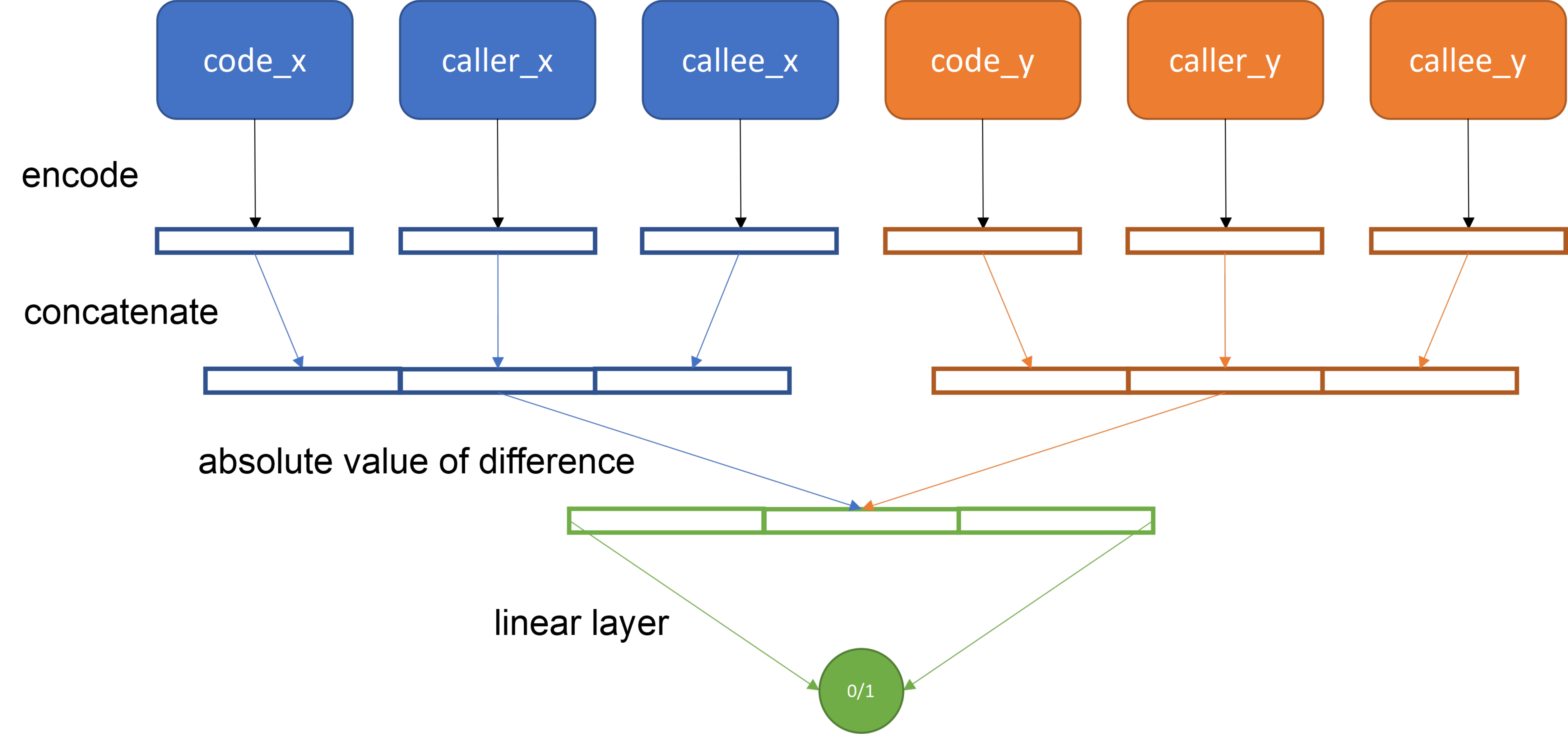}} 
    \subfloat[Max-pooling]{\includegraphics[width=0.33\linewidth]{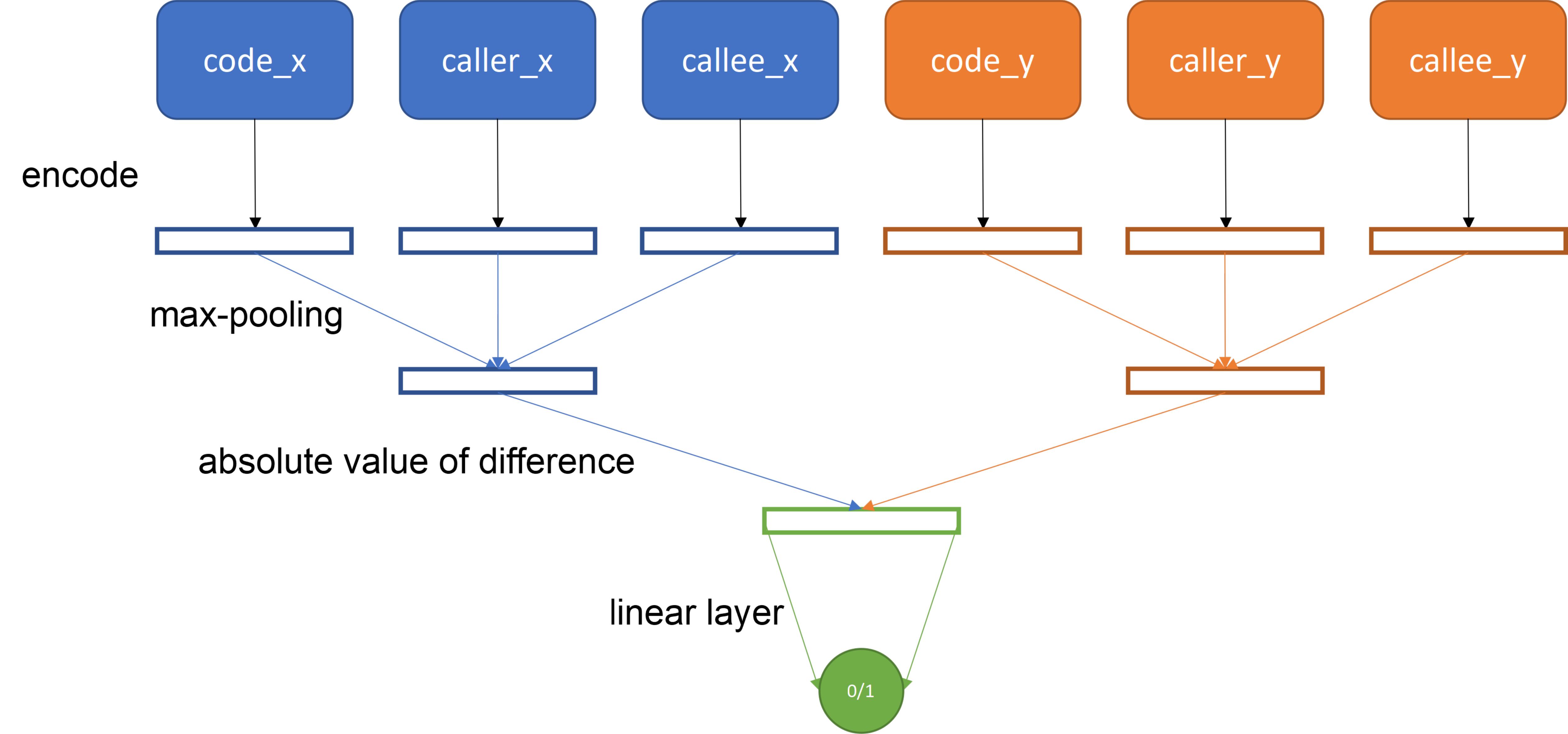}} 
    \subfloat[Concatenation of absolute difference]{\includegraphics[width=0.33\linewidth]{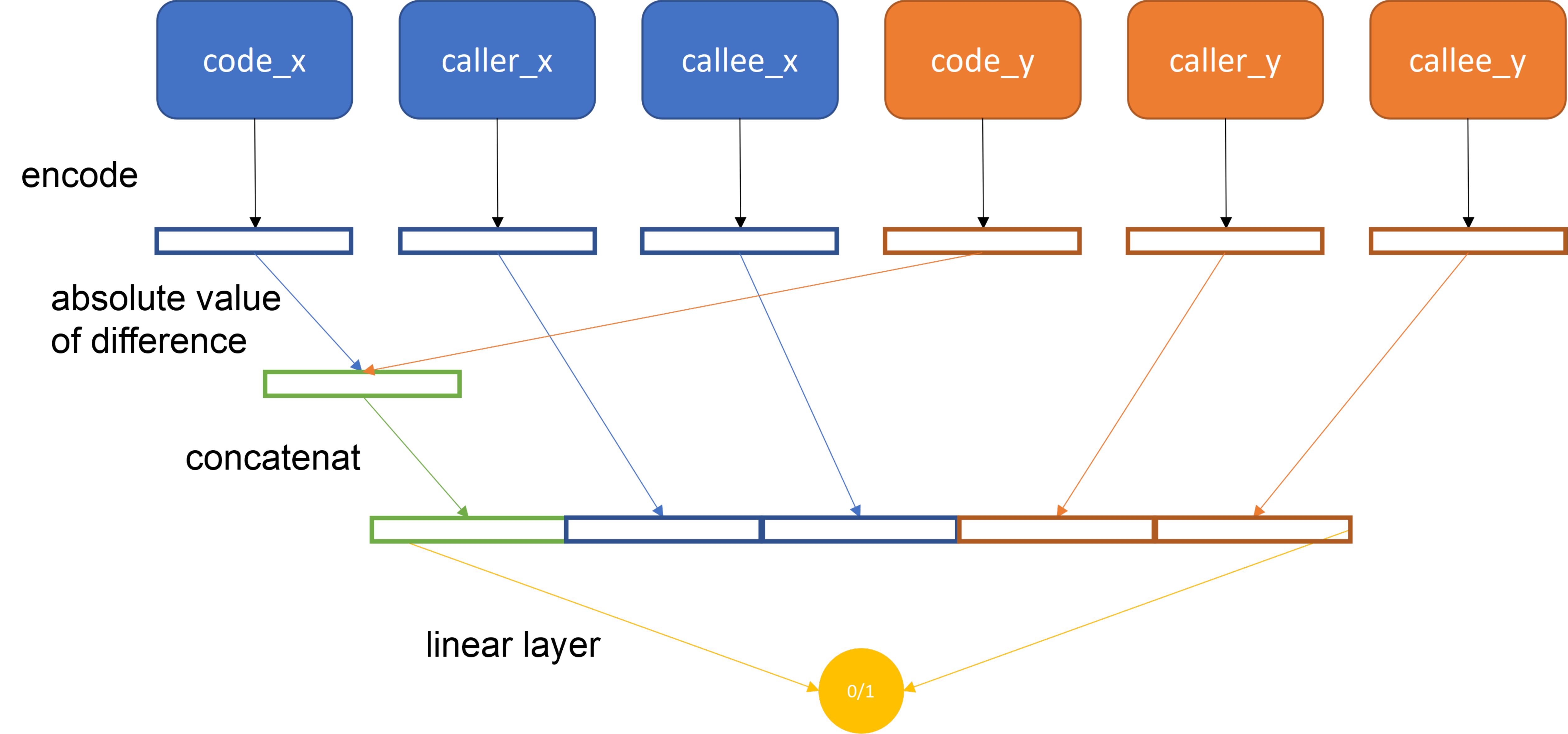}} \\
    \caption{Aggregation approaches in the clone detection scenario}
    \label{fig:clone}
\end{figure*}

ASTNN relies on 200-dimensional vectors to represent a source code snippet. Since there are multiple ways in which data can be combined in the context of machine learning (e.g., pooling~\cite{gholamalinezhad2020pooling}), we experiment with different approaches to combine the representation of a method and the representation of its context, according to the two application scenarios detailed above. Our focus in this preliminary study is not on inventing new aggregation approaches, but on evaluating well-established approaches for our problem domain.

For code clone detection, we explore three aggregation methods as shown in Figure~\ref{fig:clone}. In the \textit{concatenation} scenario, the 200-dimensional representation of a method is concatenated with the 200-dimensional representations of its caller and callee, resulting in two 600-dimensional representations for each code pair. We calculate the absolute value of the difference of these vectors, and send the result into a linear layer and a sigmoid layer to determine whether the two methods are clones of each other. In the \textit{max-pooling} scenario, we rely on pooling to select the maximum values of the method, caller, and callee vectors in each dimension to form a new vector. In the \textit{concatenation of absolute difference} scenario, we reverse the process by calculating the difference between the method vectors first and then performing the concatenation of all relevant context, i.e., callers and callees of both methods.

\begin{figure}
    \centering
    \subfloat[Concatenation]{\includegraphics[width=0.5\linewidth]{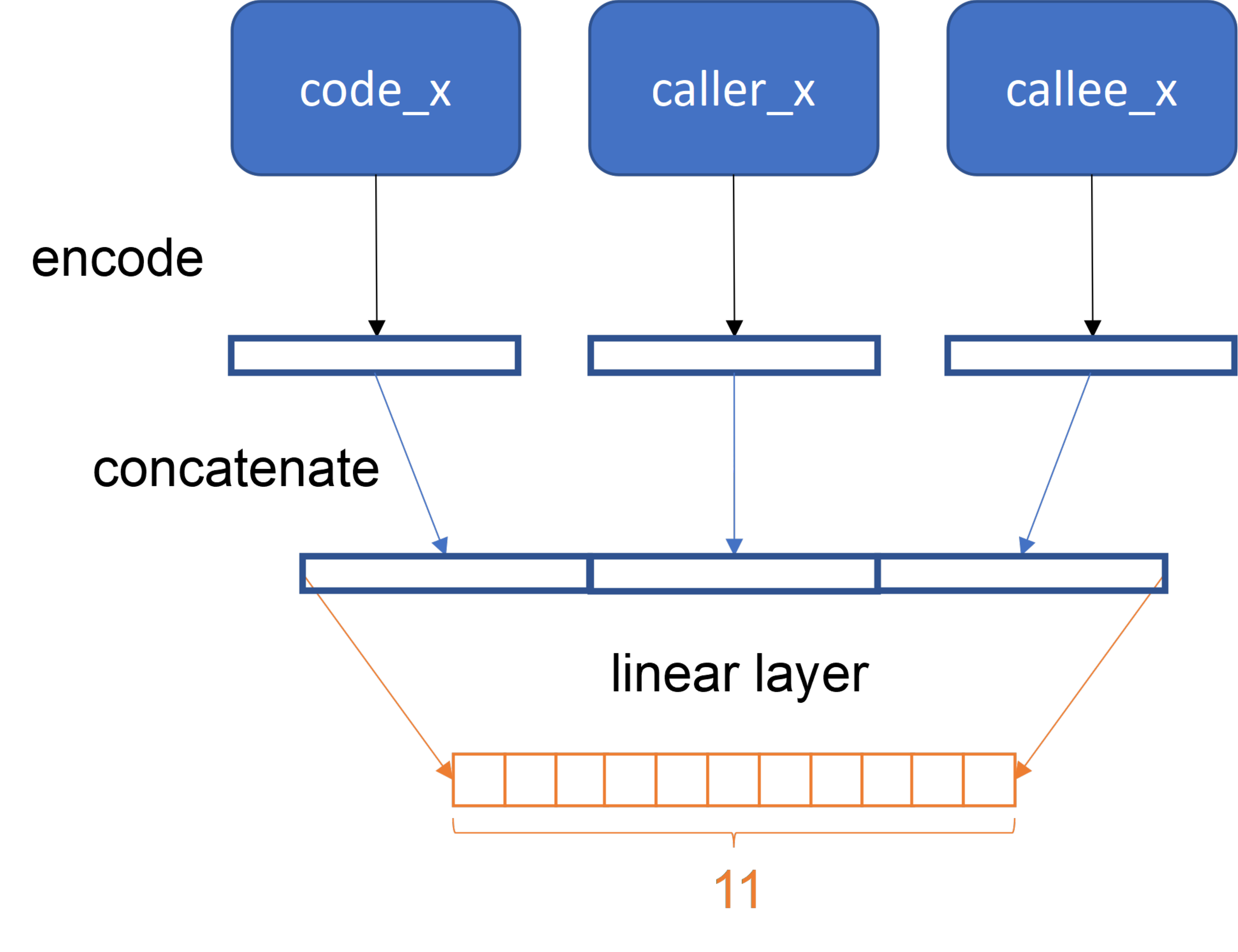}} 
    \subfloat[Max-pooling]{\includegraphics[width=0.5\linewidth]{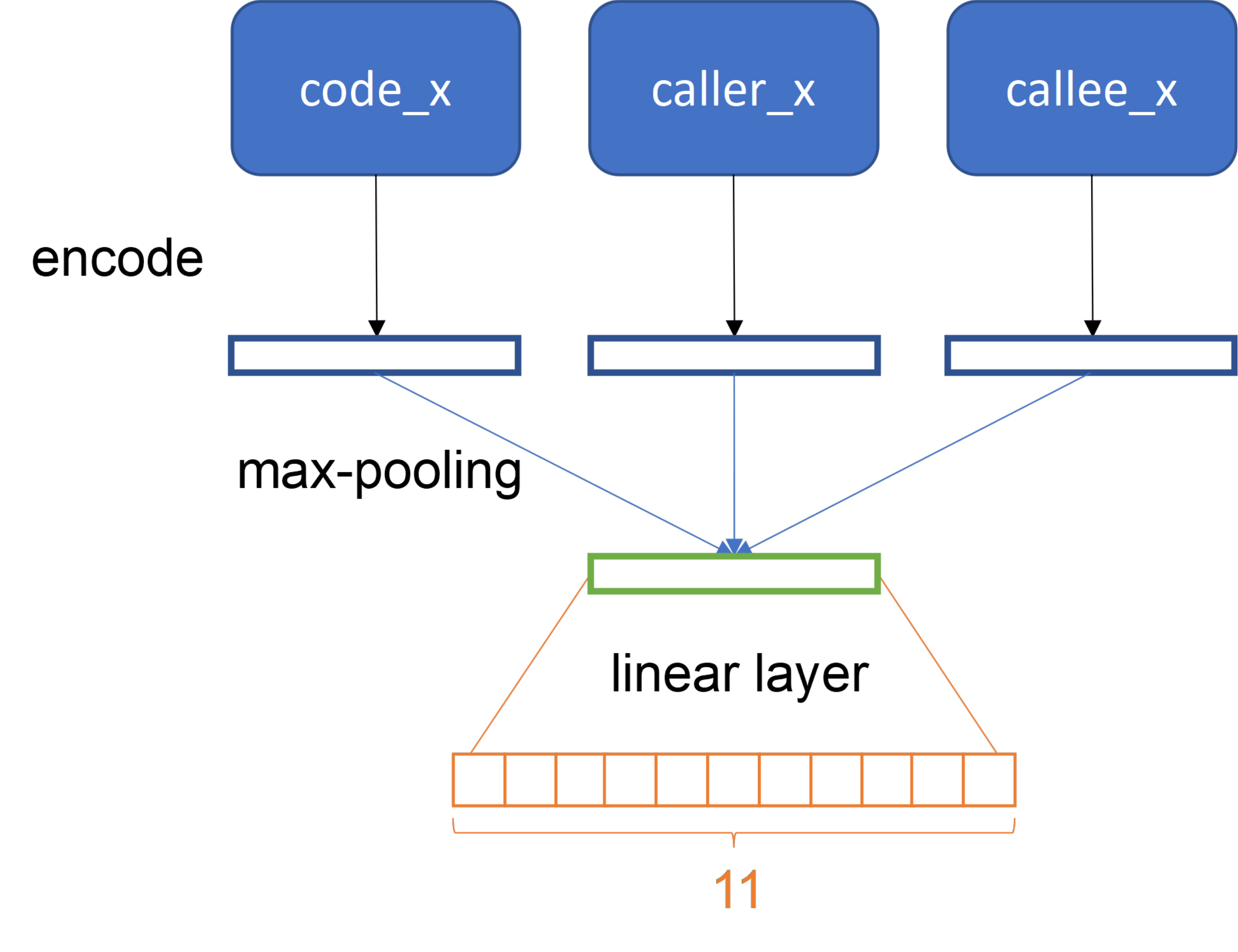}} \\
    \caption{Aggregation approaches in the classification scenario}
    \label{fig:class}
\end{figure}

For code classification, our unit of analysis are individual methods instead of method pairs. We explore two aggregation methods as shown in Figure~\ref{fig:class}. In the \textit{concatenation} scenario, we concatenate the method, caller, and callee vectors and use a softmax layer~\cite{bridle1990probabilistic} to assign each method to one of eleven classes. In the \textit{max-pooling} scenario, we use pooling to select the maximum values of the method, caller, and callee vectors in each dimension to form a new vector, which is then passed to the softmax layer.

\subsection{Results}

\begin{table}
\centering
\caption{Performance in the clone detection scenario}
\label{tab:clone}
\begin{tabular}{lrrrr}
\toprule
Method                       & Accuracy & Precision & Recall  & F1      \\
\midrule
Without Context              & 0.825 & 0.857 & 0.600 & 0.706 \\
Concatenation                & 0.807 & 0.800 & 0.600 & 0.686 \\
Max-Pooling                  & 0.789 & 0.700 & \textbf{0.700} & 0.700 \\
Difference \& Concatenation  & \textbf{0.860} & \textbf{0.929} & 0.650 & \textbf{0.765} \\
\bottomrule
\end{tabular}
\end{table}

\begin{table}[]
\centering
\caption{Performance in the classification scenario}
\label{tab:class}
\begin{tabular}{l@{\hspace{.5em}}r@{\hspace{.5em}}r@{\hspace{.5em}}r@{\hspace{.5em}}r}
\toprule
Method                              & Accuracy & Precision & Recall & Macro-F1 \\
\midrule
Without Context                     & 0.571   & 0.571    & 0.402 & 0.397   \\
Concatenation                       & \textbf{0.810}   & 0.748    & 0.686 & 0.704   \\
Max-Pooling                         & 0.790   & \textbf{0.836}    & \textbf{0.707} & \textbf{0.747}   \\
\hline
Concatenat.~w/ Random Context   & 0.771   & 0.681    & 0.637 & 0.633   \\
Max-Pooling w/ Random Context     & 0.800   & 0.747    & 0.698 & 0.711   \\
\hline
Max-Pooling w/ Caller Context     & 0.733   & 0.651    & 0.654 & 0.647   \\
Max-Pooling w/ Callee Context     & 0.676   & 0.697    & 0.530 & 0.550   \\
\bottomrule
\end{tabular}
\end{table}

Table~\ref{tab:clone} summarises our results for the code clone detection scenario. Without any customisation related to context, ASTNN achieves an F1 score of 0.706. This performance degrades when adding context via concatenation and max-pooling, with F1 scores of 0.686 and 0.700, respectively. However, we see an improvement in performance for the aggregation method of concatenating the absolute difference of the method vectors along with caller and callee context of both, for an F1 score of 0.765 (an 8\% improvement). We observe similar trends for accuracy, and note that the best recall was achieved in the max-pooling scenario.

These results show that adding additional context to source code representations for deep learning can have a positive impact on the performance of a downstream software engineering task. The way in which additional context is represented and aggregated can determine whether performance improves or decreases, so it is important to choose an approach that yields the best results.

Table~\ref{tab:class} summarises our results for the classification scenario. Without adding context, ASTNN achieves a Macro-F1 score of 0.397 on our data, across eleven classes. Adding context substantially improves this performance, with Marco-F1 scores of 0.704 and 0.747 for concatenation and max-pooling, respectively.

We note that the addition of context from the call hierarchy alone does not necessarily explain the improved performance of code classification. The baseline model without added context has to assign methods to one of eleven projects based on information from a single method, whereas in the other two scenarios (concatenation and max-pooling), the models have access to information from up to three methods from the same project. It is unsurprising that a classifier achieves better performance on this task if it has up to three data points for each decision instead of just one.

To determine the extent to which context from the call hierarchy specifically can improve performance, we compared performance against additional baselines which add random methods from same project as additional context. Table~\ref{tab:class} shows that some of the performance gain can indeed be attributed to information from the call hierarchy: Compared to adding random context, the performance improves by 11\% for concatenation (from an F1 score of 0.633 to an F1 score of 0.704) and 5\% for max-pooling (from an F1 score of 0.711 to an F1 score of 0.747). The table further shows that adding caller or callee context in isolation is not sufficient---in fact, adding callee context only led to a degradation of performance.

We answer our two research questions as follows:

\begin{mdframed}[style=mystyle,frametitle=Summary]
Our preliminary study shows that encoding additional context in source code representations for deep learning can improve the performance of a state-of-the-art deep learning model for two downstream tasks (RQ1). In terms of aggregating the representation of code and its context (RQ2), we find that the aggregation approach has a substantial impact, and performance can degrade if an inadequate aggregation approach is chosen.
\end{mdframed}

\section{Research Agenda}
\label{sec:agenda}

The preliminary study that we conducted yields encouraging results, so in this section, we outline our broader research agenda for adding context to source code representations for deep learning. We discuss different types of context, different aggregation methods, and empirical studies in the following paragraphs.

\subsection{Other Context}

When developers understand source code, they benefit from additional context, e.g., in the form of documentation~\cite{blinman2005program}, execution traces~\cite{zaidman2005applying}, or navigation patterns~\cite{deline2005easing}. We argue that source code representations can benefit from such context as well. We distinguish two types of context:

\begin{itemize}

\item Definite Context: The identification of definite context is based on facts and does not rely on probabilistic reasoning. Types of definite context for a code fragment include its version history, its execution traces, and its call hierarchy.

\item Possible Context: The identification of possible context is affected by uncertainty, similar to how a developer might search for information about a code fragment but be faced with uncertainty as to whether a particular piece of information, such as a Stack Overflow thread, is actually related to this code fragment. Types of possible context for a code fragment include its documentation, issues that have been reported against it, and its rationale. While research has made great progress towards establishing such traceability links~\cite{cleland2014software}, information inference often happens under uncertainty~\cite{robillard2017demand}.

\end{itemize}

We argue that source code representations could potentially benefit from all of these types of context, depending on encoding and downstream tasks.

\subsection{Aggregation}

Pooling is an aggregation technique used in deep learning to reduce the number of parameters and improve performance. It works by combining multiple input features into a single feature, which is then passed through the network. This approach can be used with convolutional neural networks (CNNs) and fully connected nets, and can be applied at different levels in the network depending on what is being optimised. An ideal pooling method is expected to extract only useful information and discard
irrelevant details~\cite{gholamalinezhad2020pooling}. We argue for the use of two classes of aggregation techniques:

\begin{itemize}

\item General-purpose pooling: In addition to max-pooling and concatenation used in our preliminary study, many other pooling techniques have been proposed in the context of CNNs, e.g., average pooling, stochastic pooling, and weighted pooling. We refer readers to Gholamalinezhad and Khosravi~\cite{gholamalinezhad2020pooling} for an overview.

\item Domain-specific aggregation: The most suitable way of aggregating data will often depend on the domain and the downstream task, just as in our preliminary study where the aggregation technique which calculates the difference between two method representations achieved the best performance for code clone detection. We expect to see similar advantages when exploring other domain-specific aggregation methods, e.g., by taking time series information into account when aggregating version histories or by encoding probabilities when context inference was done under uncertainty.

\end{itemize}

We argue for extensive experiments to evaluate the effect of these techniques on aggregating the representation of source code and its context for deep learning, as well as the exploration of different neural network architectures and their suitability for this task.

\subsection{Dissection}

Interpreting and dissecting what a deep learning model has learned has become a key ingredient for the validation of such models~\cite{montavon2018methods}. We posit that developers and tool builders can benefit from understanding which context a model found beneficial for which software engineering task and why, with particular focus on the interplay of source code and its context. Ultimately, we argue for a feedback loop: By enabling deep learning methods to benefit from the same context that developers have access to, we can improve the models' understanding of source code, and by dissecting what the models have learned~\cite{samek2016evaluating}, we can improve how developers complete their tasks, benefiting from knowledge about how to best make use of source code and its context.

\bibliographystyle{IEEEtran}

\end{sloppy}
\end{document}